# An optimized TEM specimen preparation method of quantum nanostructures

*Hongguang Wang\*, Vesna Srot, Bernhard Fenk, Gennadii Laskin, Jochen Mannhart,*

*Peter A. van Aken*

Max Planck Institute for Solid State Research, 70569 Stuttgart, Germany

\* hgwang@fkf.mpg.de

**Abstract**

Electron transparent TEM lamella with unaltered microstructure and chemistry is the prerequisite for successful TEM explorations. Currently, TEM specimen preparation of quantum nanostructures, such as quantum dots (QDs), remains a challenge. In this work, we optimize the sample-preparation routine for achieving high-quality TEM specimens consisting of $SrRuO_3$ (SRO) QDs grown on $SrTiO_3$ (STO) substrates. We demonstrate that a combination of ion-beam-milling techniques can produce higher-quality specimens of quantum nanostructures compared to TEM specimens prepared by a combination of tripod polishing followed by $Ar^+$ ion milling. In the proposed method, simultaneous imaging in a focused ion-beam device enables accurate positioning of the QD regions and assures the presence of dots in the thin lamella by cutting the sample inclined by 5° relative to the dots array. Furthermore, the preparation of TEM lamellae with several large electron-transparent regions that are separated by thicker walls effectively reduces the bending of the specimen and offers broad thin areas. The final use of a NanoMill efficiently removes the amorphous layer without introducing any additional damage.

Keywords: Quantum nanostructures, Focused ion beam, Tripod polishing, Ion milling, NanoMill

**Introduction**

Quantum matter nanostructures offer possibilities for artificial manipulation of novel electron systems (Boschker and Mannhart, 2017; Harrison and Valavanis, 2016), which have been playing an important role in applications like quantum computing and sensing (Senellart et al., 2017). Among them, zero-dimensional electron systems, also known as quantum dots (QDs), attract great attention due to their size confinement (Alivisatos, 1996; Mannhart et al., 2016), which splits the electron energy levels and generates intriguing quantum phenomena. In our previous work, we showed that the magnetic Curie temperature of SRO QD nanostructures increases if the dot size is reduced to 30 nm (Laskin et al., 2019). Atomically resolved investigations of their structure and chemistry are crucial for the understanding of their fundamental physical mechanisms.

A transmission electron microscope (TEM) is a superior tool used for nanoscale characterization of QDs. Particularly after the invention of aberration correctors, TEM enables the observation of microstructure, electronic structure, and local chemistry at the atomic scale (Pennycook and Nellist, 2015). However, the quality of TEM specimens is often a major limiting factor (David B. Williams, 2009; Srot et al., 2014). Uniformly thin TEM specimens with a clean surface and a thickness of up to 50 nm are prerequisites for TEM observations (Brydson, 2011). Due to the small size of QDs (down to 15 nm) (Laskin et al., 2019; Wolfsteller et al., 2010), it is particularly difficult to obtain electron transparent lamellae, where QDs have been cut through their central part. TEM investigations of 80 nm SRO QDs were reported (Ruzmetov et al., 2005), displaying an extensive amorphous layer covering $SrRuO_3$ (SRO) QDs and a noticeable shape change caused by the TEM specimen preparation procedure. Due to technological advancements, the size of QDs has been decreasing over the years, making TEM specimen preparation even more challenging and demanding.

In this work, by utilizing aberration-corrected (scanning) transmission electron microscopy ((S)TEM), we compare the quality of TEM specimens that were prepared by (i) tripod polishing followed by Ar$^+$ ion milling (TP&IM) and (ii) focused ion beam cutting with subsequent low-energy milling and cleaning in a Fischione NanoMill (FIB&NM).

**Experimental methods**

> *Sample preparation*

SRO quantum-dot nanostructures were obtained by patterning an epitaxial SRO thin film grown on a STO substrate by electron beam lithography (JEOL JBX6300) with an electron-beam energy of 100 keV. The dot size and spacing are controlled during the patterning process. Figure 1(a) shows a sketch of the SRO QD nanostructures. Scanning electron microscopy (SEM) images of the patterned SRO QDs on the STO substrates are shown in Fig. 1(b, c). Here, the SRO QDs with 30 nm diameter are periodically arranged with a spacing of about 100 nm.

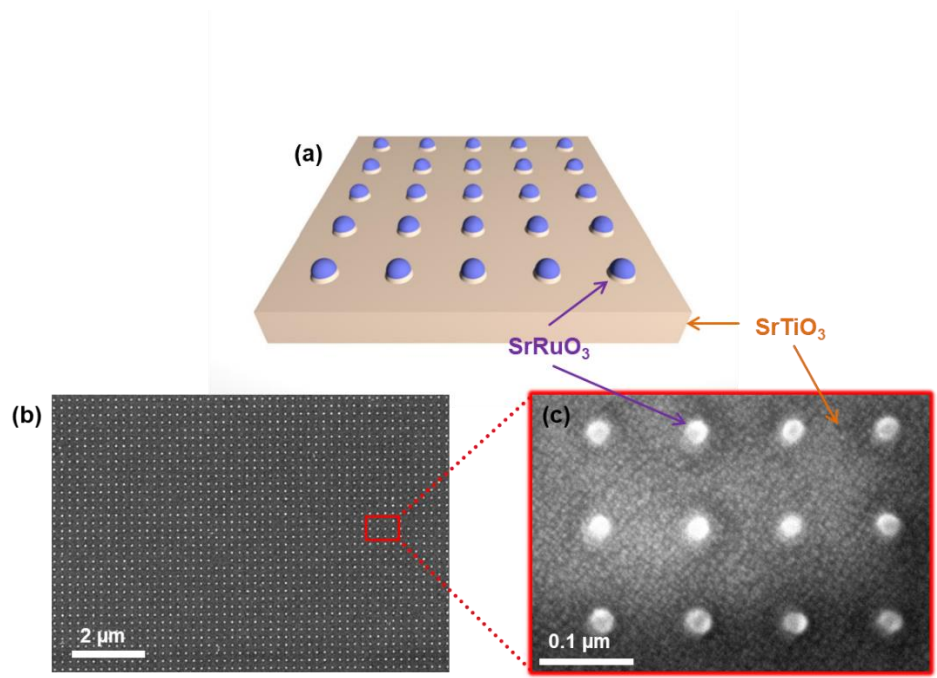

Figure 1. (a) Sketch of a SRO QD array on a STO substrate. (b) SEM image of the SRO QDs nanostructure. (c) Magnified SEM image of the region within the rectangle marked in (b).

- *TEM specimen preparation*

In this work, we have prepared electron-transparent TEM lamellae by two combinations of two different methods, tripod polishing followed by $Ar^+$ ion milling (TP&IM) and focused ion beam cutting with subsequent low-energy milling and cleaning in a Fischione NanoMill (FIB&NM). All prepared cross-sectional samples were observed by (S)TEM along the [100] direction of the STO substrate.

*(i) Tripod polishing with subsequent $Ar^+$-ion milling*

Samples were cut into slabs of 0.5 mm width and 1.5 mm length with a wire saw (Well, Model 3242), as shown in Figs. 2(a-c). A thin layer of glass was mounted to the surface of the slab with glue (Epoxy bond 110) to protect the surface structure from contamination and damage during further processing. Then, the samples were attached to a Pyrex specimen holder using Crystal Bond thermoplastic wax for mechanical polishing. An Allied MultiPrep System was used for automated tripod polishing of the samples in cross-sectional geometry (Srot et al., 2014; Voyles et al., 2003). Diamond lapping films (DLF) with grain sizes of 3.0, 1.0, 0.5, and 0.1 µm were used. Firstly, one side was flatly polished until the sample thickness was below 250 µm. During the polishing of the second side, a wedge angle of 1° was introduced as the thin part of the wedge was below 70 µm. Finally, the thin part of the wedge was polished down to about 10 µm using a 0.1 µm DLF (Fig. 2(d)). The polished samples were removed from the holder and glued to a molybdenum half-ring with M-Bond 610 epoxy (Fig. 2(e)). During the mechanical polishing process, the sample thickness was regularly checked with an optical microscope. A Gatan Precision Ion Polishing System (PIPS II, Model 695) was used for $Ar^+$ ion milling of the samples at liquid $N_2$ temperature. To thin the sample and to reduce the ion-beam-induced damage, the acceleration voltage during $Ar^+$ ion milling was progressively lowered from 3.3 kV to 2.0 kV and 0.3 kV. The central region of the thin part of the wedge was milled using the ion beam at an angle of 8°. An optical microscopy image (Zeiss AxioCam HRc) of the final TEM specimen is shown in Fig. 2(f). Optical interference fringes indicate that the thickness close to the curved region is thin enough for TEM observation (Hudson et al., 2004). The edges marked by arrows in Fig. 2(f) are the regions of interest, where (S)TEM investigations were performed.

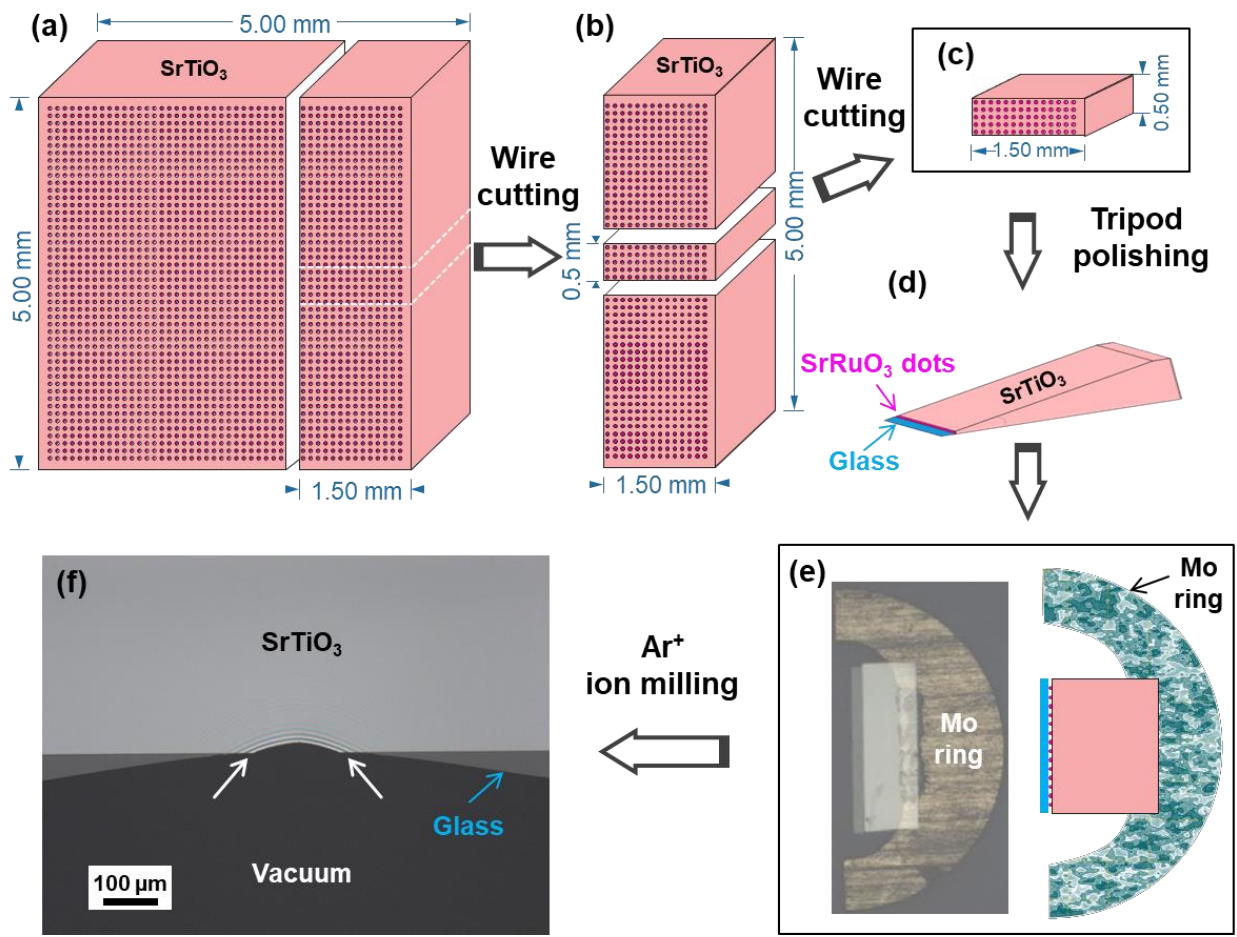

Figure 2. Schematic diagram of TEM sample preparation by tripod polishing followed by $Ar^+$ ion milling (TP&IM). (a–c) Cutting slabs of 1.5 mm width and 0.5 mm thickness using a wire saw. (d) Preparation of wedge-shaped specimen by tripod polishing. (e) Attaching a tripod-polished sample

to a molybdenum half ring. (f) Milling the central region of the thin part of the wedge by $Ar^+$ ion milling to obtain the finalized TEM specimen. The white arrows mark the areas of interest; the blue arrow marks the glass layer.

**(ii)** *Focused ion beam cutting followed by $Ar^+$ ion thinning and polishing in a Fischione NanoMill system*

The focused ion beam system (Zeiss CrossBeam XB 1540) makes use of a combination of a focused ion beam (FIB) and an electron beam for scanning electron microscopy (SEM). This combination allows for site-specific preparation of the samples (Mayer et al., 2011; Wang, 2013) and thus is ideally suited for TEM specimen preparation of quantum nanostructures. To ensure the presence of central-cut SRO QDs in the electron transparent lamellae, the cutting direction needs to be precisely determined. The spacing between the dots is 100 nm, the QD size is 30 nm, and the final sample thickness is ca. 20 nm. Since the width of every thinned region in the lamellae will be around 2000 nm, the cutting angle for our sample should be between arctan(130nm/2000nm) = 3.7° and arcsin(25nm/130nm) = 11.1° for a dot size of 30 nm in order to have at least one central-cut dot for TEM observations along the [100] direction. Here, we used a cutting angle of 5° between the cutting direction and the parallel rows of QDs.

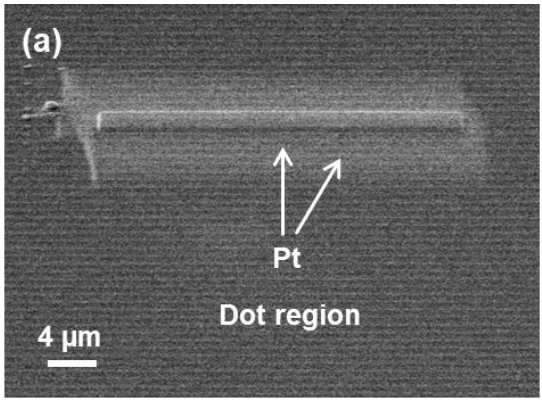
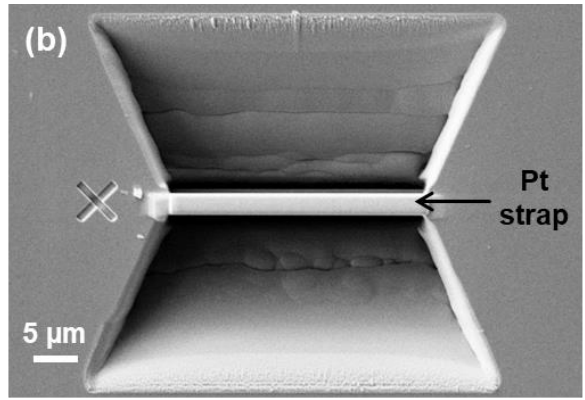
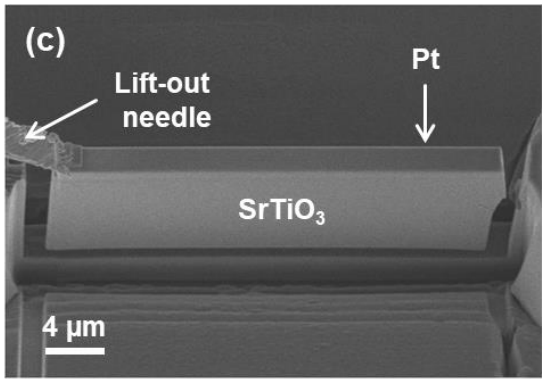
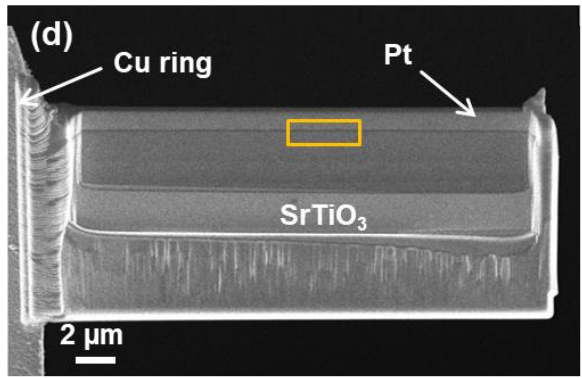
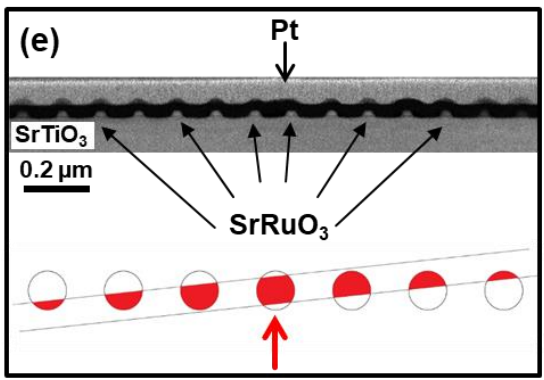
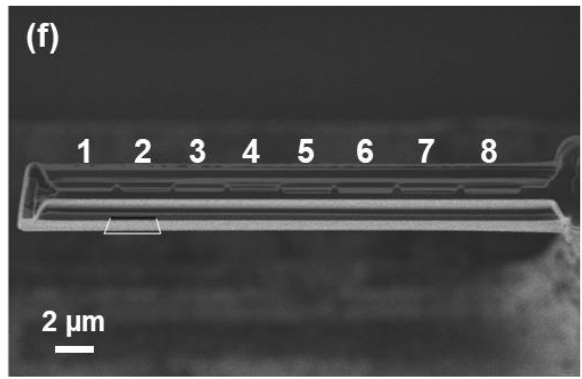
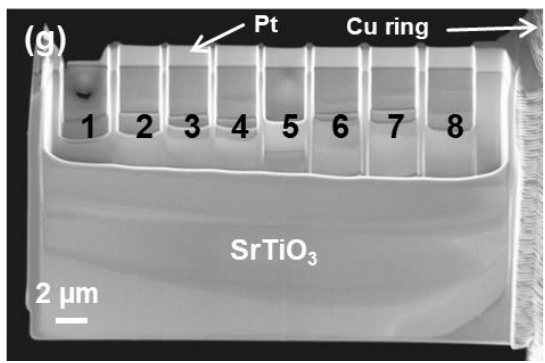
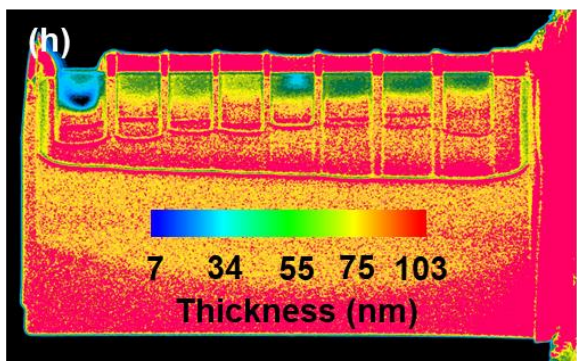

Figure 3. TEM sample preparation with a focused ion beam (FIB). (a) Deposition of a Pt protective layer on the bulk surface via electron-beam and ion-beam induced deposition. The straight Pt strap is deposited at an angle of 5° with respect to the QD rows. (b) Cutting of two trenches on both sides of the Pt strap with a $Ga^+$ ion beam leaving a thin section of material isolated at the center. (c) Separation of the lamella from the bulk with an U-shaped undercut by an in-situ lift-out of the specimen with a micromanipulator (Kleindiek MM3A). (d) Attaching of the FIB lamella to a half-moon-shaped copper (Cu) grid (Omniprobe 3 post lift-out grid) using Pt. (e) Upper part: Imaging of the QDs by high-angle annular dark-field scanning transmission electron microscopy (HAADF-STEM) of the yellow box region marked in (d). Bottom part: A sketch of the cutting geometry of QDs at different positions. (f) Thinning of the FIB lamella from two sides (top view). Eight regions of interest were set for fine thinning. (g) Backscattered electron image of the final FIB lamella, that was thinned from both sides. 8 regions are separated by thicker walls. (h) Thickness map of the lamella based on the backscattered electron signal, where the STO substrate was used as a reference for the thickness evaluation.

Before FIB preparation, we pretreated the SRO QD nanostructure. The surface of the sample was firstly wiped by an organic solvent (acetone) to clean possible surface contaminations, which is necessary to reduce artifacts of preferential milling (Langford and Petford-Long, 2001). Then, the sample was coated with a 30 nm thick carbon layer by thermal evaporation (Leica EM ACE 600 coater system) to protect the surface of the sample and to make it conductive for reducing charging effects (Schaffer et al., 2012). Next, we deposited a Pt protection layer by electron-beam induced deposition (EBID) and ion-beam induced deposition (IBID), successively. This protection layer with a size of 30 ☐m x 2 ☐m is inclined by 5° with respect to the SRO QD rows. As shown in Fig. 3(a), a Pt strap was deposited on the surface of the sample. Two trenches on both sides of the Pt strap were milled with the $Ga^+$ ion beam at 30 kV and 10 nA followed by milling at 30 kV and 2 nA (Fig. 3(b)). A U-shaped undercut separated the lamella from the bulk. Thereafter, the *in-situ* micromanipulator needle (Kleindiek MM3A) was attached to the edge of the lamella (Fig. 3(c)), and the sample was lifted out and fixed to a a half-moon-shaped copper (Cu) grid using Pt. The

SEM image of the obtained FIB lamella is shown in Fig. 3(d). The geometry of QDs in the obtained FIB lamella appears different due to the applied cutting angle, as shown in the ADF-STEM image and the sketch in Fig. 3(e). According to the gradual variation of the geometry of the QDs in this row, one can easily find the central-cut QD, which periodically appears in different areas of the FIB lamella. Red-colored regions (the sketch in Fig. 3e) indicate the remaining part of QDs. This FIB lamella was additionally milled by $Ga^+$ ion from both sides (Fig. 3(f-g)). Then FIB milling was performed at an acceleration voltage of 30 kV by using a beam current of 200 pA for coarse milling, which was gradually reduced to 50 pA for fine thinning with the milling angle reduced from 2° to 1.2°. Eight regions were set for fine thining and thick walls lie in-between, which is beneficial to minimize bending effects during TEM characterization. Among them, region 1 was further milled at an acceleration voltage of 3.0 kV using a beam current of 50 pA at a milling angle of 5°. The thickness map of the finalized FIB lamella (Fig. 3(h)) was obtained by the Zeiss SmartEPD software package via analyzing the backscattered intensity using STO as a reference (Drouin et al., 2007; Salzer et al., 2009). The thickness of region 1 is about 30 nm. The thickness of regions 2 to 8 ranges from 50 nm to 70 nm. Region 2 to 8 were additionally thinned and cleaned using a Fischione Model 1040 NanoMill TEM specimen preparation system (E.A. Fischione Instruments, Inc.) at liquid nitrogen temperature. This sample treatment using the NanoMill is ideal for post-FIB processing (Cerchiara et al., 2011; Dienstleder et al., 2017; Giannuzzi and Stevie, 1999; Nowakowski et al., 2017), since it efficiently removes amorphized material and implanted $Ga^+$ ions caused by FIB preparation using a highly focused $Ar^+$ ion beam (as small as 1 µm in diameter). With the Nanomill, the FIB lamella was milled from both sides, starting with an ion-beam acceleration voltage of 900 V, which was gradually reduced to 700 V for thinning and finally to 400 V for cleaning with a beam current starting from 100 pA and being reduced to 50 pA.

In this work, we employed three different ion-milling machines (NM, FIB, and PIPS). The experimental conditions, including the ion energy, the angle between the ion beam and specimen, the direction of the ion beam, the ion current density, and the specimen temperature, determine the ion-beam-induced damage. The resulting details of our experiments are displayed in Table 1. Their effects on the quality of the final TEM specimen will be discussed in the following parts.

|  | PIPS | FIB | NM |
| --- | --- | --- | --- |
| Ion beam | Ar$^+$ | Ga$^+$ | Ar$^+$ |
| Energy (kV) | 3.3 / 2.0 / 0.3 | 30 / 3 | 0.9 / 0.7 / 0.4 |
| Angle (°) | 8 | ±(2-1.2) (30 kV) / ±5 (3 kV) | ±10 |
| Direction of the ion beam | S-D | D-S | D-S |
| Beam current density | ~10 mA/cm$^2$ | ~(5-16) A/cm$^2$ | up to 1 mA/cm$^2$ |
| Cooling condition | Liquid-N$_2$ | No cooling | Liquid-N$_2$ |

Table 1. Experimental details for the different ion milling processes. D (dot) – S (substrate) represents the ion beam direction from the dot to the substrate and vice versa. The current density (current per unit area) of the electron beam has been calculated using the formula $I/\pi(d/2)^2$, where $I$ is the beam current and $d$ is the beam diameter.

➢ *TEM characterization*

High-resolution TEM (HRTEM) imaging was performed at 200 kV with a JEOL ARM200F (JEOL Ltd.) microscope equipped with a CETCOR image corrector (CEOS GmbH). HAADF-STEM imaging was carried out at 200 kV with a JEOL ARM200F (JEOL Ltd.) equipped with a DCOR probe corrector (CEOS GmbH). Both microscopes are equipped with a cold-field emission gun (CFEG) and a Gatan GIF Quantum ERS imaging filter (Gatan Inc. Pleasanton, USA) with dual-EELS acquisition capability. For STEM imaging, the microscope was operated at a convergence semi-angle of 20.4 mrad, resulting in a probe size of 0.8 Å. Collection angles of 70−300 mrad were used to obtain the HAADF-STEM images. Energy-dispersive X-ray spectra (EDX) were obtained by acquiring area scans using a 100 mm$^2$ JEOL Centurio SDD-EDX detector and the Thermo Noran System 7 EDX system (Thermo Fischer Scientific Inc.). Relative thickness ($t/\lambda$) measurements were performed by acquiring low-loss EEL spectra at a dispersion of 0.25 eV/channel using the routine implemented in Digital Micrograph (Gatan), where $t$ is the absolute sample thickness and $\lambda$ the inelastic mean free path.

**Results and discussions**

Fig. 4(a) displays the low-magnification HAADF-STEM image of SRO QDs on a STO substrate acquired from the TEM specimen prepared by TP&IM, showing a regular arrangement of QDs. The HAADF-STEM image of a single SRO QD is shown in Fig. 4(b). It can be seen that an amorphous layer is present on the top of the dot, which may result from a surface damage layer induced by the patterning process. EDX measurements performed from 3 regions marked in Fig. 4(b) are shown in Fig. 4(c). The chemical composition measured from regions 1 and 2 is from the STO substrate. Since the signal intensity of the HAADF image in the same material is proportional to the thickness, the difference in image contrast between these two areas is attributed to the thickness differences of the TEM specimen. Since the etching process during the QD preparation extends to the STO substrate, either the TEM specimen thickness is larger than the dot size (30 nm) or the thickness difference between region 1 and region 2 originates from a non-central cutting of the dot. As expected, in the QD (position 3) only the elements Sr, Ru, and O are detected.

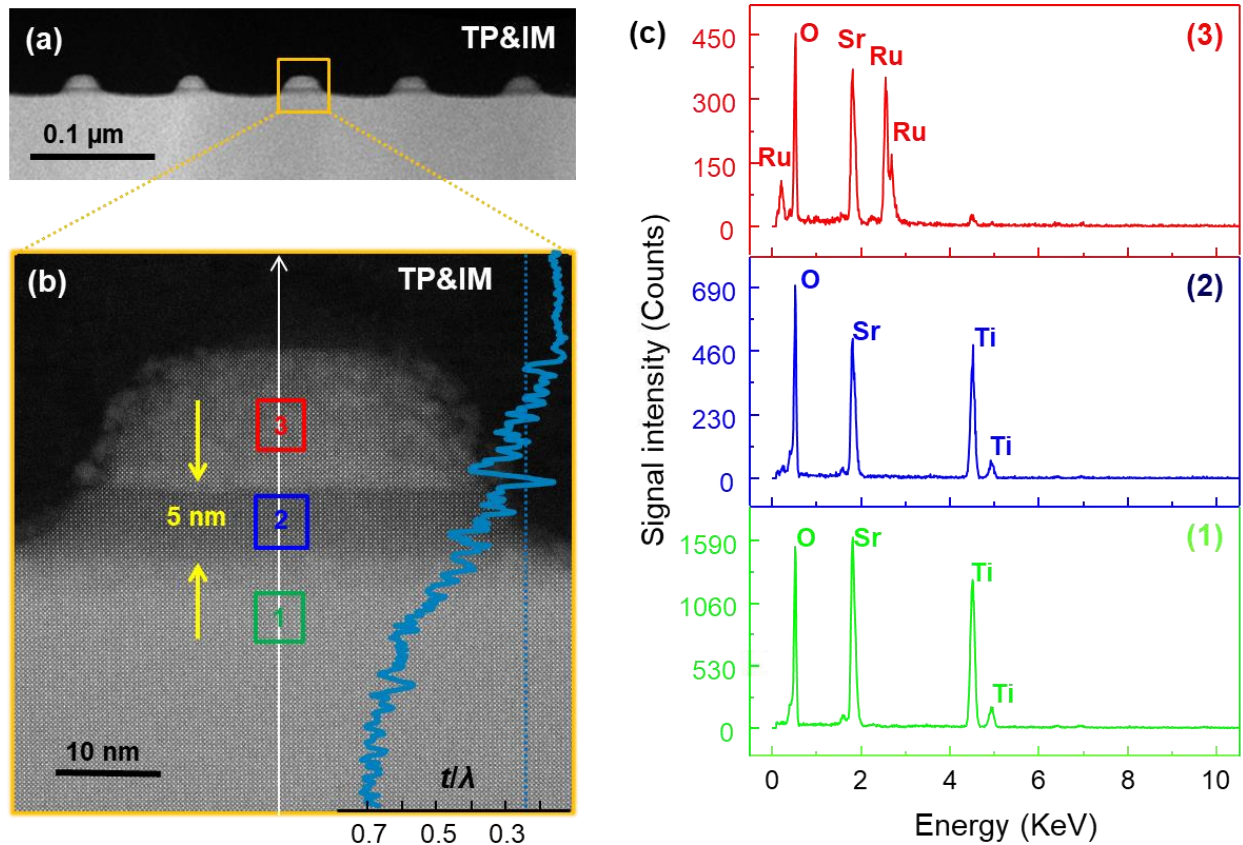

Figure 4. (a) Low-magnification HAADF-STEM image of the nanostructure cross-section prepared by tripod polishing followed by Ar$^+$ ion milling. (b) HAADF-STEM image of an individual SRO QD with a relative thickness of $t/\lambda \approx 0.25$. The superimposed profile represents the line profile of the relative thickness $t/\lambda$ across the center of the dot along growth direction as indicated with the white line. (c) Corresponding EDX spectra measured from regions (1-3) marked in (b).

For the TEM sample preparation of QD nanostructures, TP&IM has certain technical drawbacks that cannot be overcome:

a) Only a limited number of QDs are present in areas that are thin enough for TEM observations. Therefore, the possibility of finding a complete QD structure that was sliced through its center is significantly reduced.

b) Since the QDs cannot be located during the TEM sample preparation process, their presence in the final TEM lamellae is by mere chance.

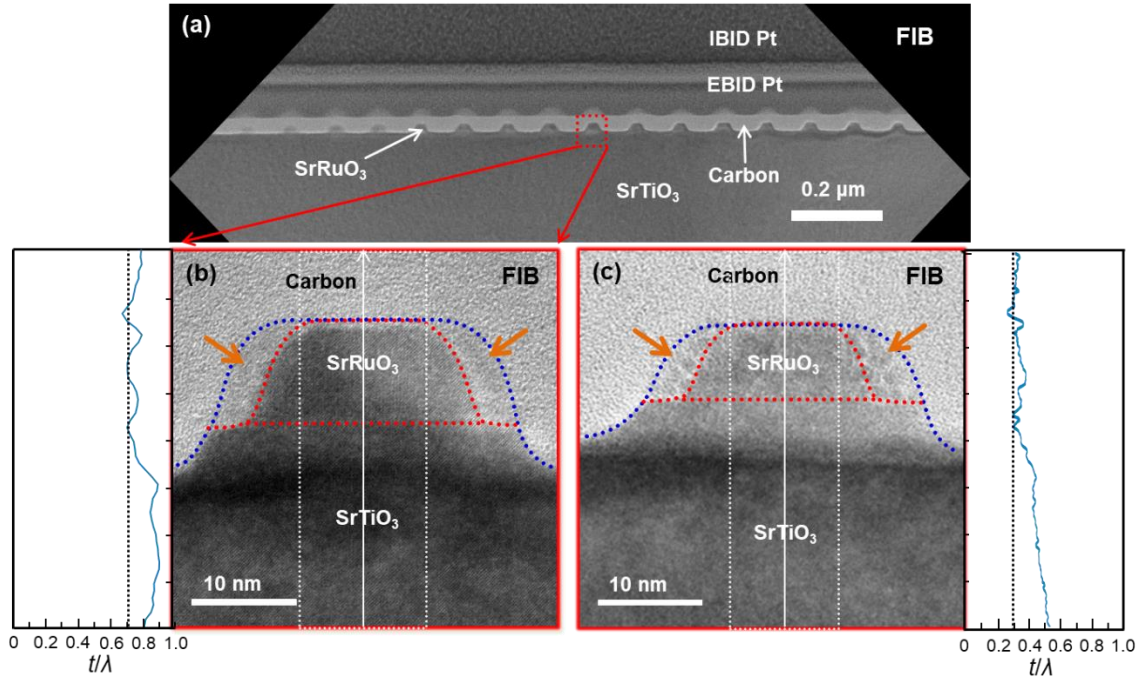

Figure 5. (a) Low-magnification BF-TEM image of a thick region ($t/\lambda \approx 0.7$) in a FIB-prepared lamella. (b), (c) are close-up images of a single SRO QD acquired from thick ($t/\lambda \approx 0.7$) and thin regions ($t/\lambda \approx 0.3$) in a FIB-prepared lamella, respectively. The line profiles on the left and right sides correspond to averaged intensity profiles of the relative thickness $t/\lambda$ across the center of the dots along the epitaxial direction as indicated in (b) and (c). Red dotted curves circle the regions of SRO QD and the blue dotted curves circle the SRO QDs surrounded by damaged SRO material (marked by orange arrows).

Fig. 5(a). shows a bright-field (BF)-TEM image of a thick region (region 2 to region 8 in Fig. 3(g)) of the FIB lamella. The close-up image of a single SRO QD is displayed in Fig. 5(b). Shadows appear at the right and left sides of the SRO QD. One area of the FIB lamella (marked as region 1 in Fig. 3(g)) was further thinned by $Ga^+$ ions to obtain a thickness below 30 nm. The TEM image

of a single SRO QD acquired from the thin region is shown in Fig. 5(c). It can be seen that shadows are still present, which are in similar shape as the shadow in Fig. 5(b). As shown in Fig. 5(b) and Fig. 5(c), both QDs are inside the carbon protection layer, revealing that shadows mostly result from the patterning process during the fabrication of QD nanostructures.

We employed a NanoMill to further thin and fine polish the FIB-prepared lamella. The measured relative sample thickness $t/\lambda$ of the Nanomill-processed FIB lamella is 0.25, which is similar to the thickness of the specimen prepared by TP&IM. The estimated mean free path $\lambda$ of SRO is about 80 nm (Iakoubovskii et al., 2008; Malis et al., 1988), giving an absolute thickness ($t$) of SRO of around 20 nm. With the combination of FIB and NM, we can easily find the locations of the central-cut dots and produce high-quality TEM specimen for high-resolution STEM studies with high accuracy and high precision.

Next, we compare the quality of TEM specimens obtained with different TEM specimen preparation methods using atomically resolved HAADF-STEM images, as shown in Figs. 6(a) and 6(b). As discussed before, the different contrast in the STO region of Fig. 6(a) is attributed to a thickness step in the STO substrate during specimen prepared by TP&IM (Fig. 6(a)). Such a step is absent for the QD cut through its center, which has been prepared by FIB&NM, as shown in Fig. 6(b). Furthermore, the signal intensity of the STO substrate in Fig. 6(b) is more homogeneous than in Fig. 6(a). In order to compare the quality of TEM specimens, we normalize the signal intensities of two images according to the formula: $I_{normalized}=2\cdot(I–I_{mean})/(I_{max}–I_{min})$, where $I_{mean}$ is the mean value, $I_{max}$ is the maximum and $I_{min}$ is the minimum value of the signal intensity. Using this formula, the HAADF signal is normalized between 1 and -1. The quality of the TEM specimen can be evaluated by observing the signal oscillation around 0 in the normalized signal intensity profiles. Fig. 6(c) and Fig. 6(d) represent the normalized intensity profiles of a region in the STO substrate

(12 nm to the interface) marked with the red box in Fig. 6(a) and 6(b). We observe that the signal oscillations in Fig. 6(d) are more uniform than those of Fig 6(c). The average maximum values are 1.00±0.08 and 1.00±0.04 for Fig. 6(c) and Fig. 6(d), respectively. Besides, the corresponding minimum is -0.61±0.08 and -0.69±0.03. In contrast to Fig. 6(c), the standard deviation in Fig. 6(d) is smaller, revealing that the TEM specimen prepared by FIB&NM has a better quality compared to that by TP&IM. The SRO part of Fig. 6(b) appears cleaner without obviously damaged regions compared to Fig. 6(a). Fig. 6(e) and 6(f) correspond to the normalized intensity profiles extracted from a region (4 nm to the interface) marked with blue boxes in Fig. 6(a) and 6(b), respectively. The averaged maximum for Fig. 6(e) and Fig. 6(f) are 0.87±0.11 and 0.90±0.08, respectively. Moreover, the corresponding averaged minimum is -0.52±0.12 and -0.45±0.10. In contrast to Fig. 6(e), the standard deviation of the signal intensity in Fig. 6(f) is smaller, revealing a better specimen quality in the SRO part. The proposed sample preparation by FIB&NM enables us to find the central-cut dots in the FIB lamella and thus minimizes the influence of the dead surface layer on the image contrast during the TEM characterization. The NanoMill can substantially reduce the beam damage and produce a sample with clean surfaces using a focussed argon-ion beam with a low beam-current density (Table 1). Based on the presented data, we have demonstrated that the optimized procedures through a combination of FIB and NM can indeed produce high-quality TEM specimens of QD nanostructures.

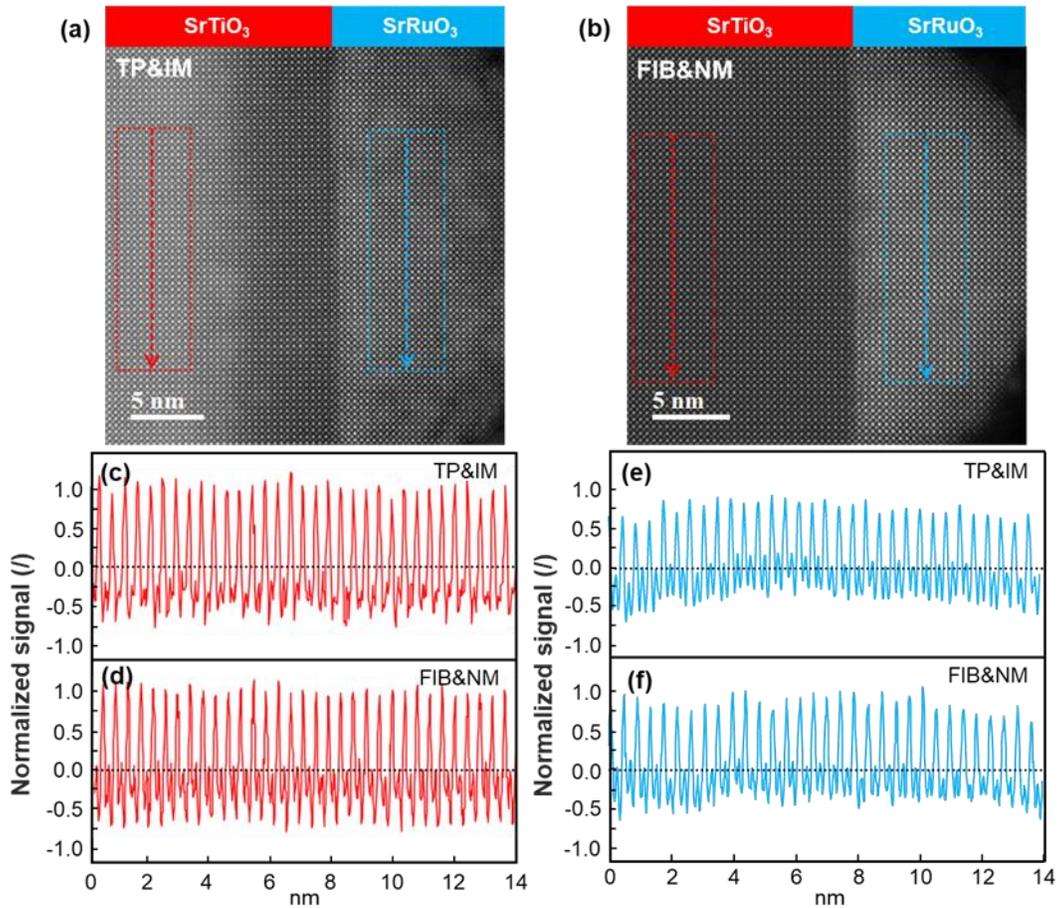

Figure 6. HAADF-STEM images of individual SRO QDs prepared by (a) tripod polishing followed by argon ion milling (TP&IM) and by (b) focused ion beam preparation followed by milling with a NanoMill (FIB&NM). The corresponding normalized intensity profiles of a region (14 nm ×5 nm) of STO substrates (12 nm to the interface) marked with red boxes in (a, b) are shown in ((c) TP&IM, (d) FIB&NM). The corresponding normalized intensity profiles ((e) TP&IM, (f) FIB&NM) of the SRO region (14 nm ×5 nm) located at 4 nm to the interface marked with blue boxes in (a, b).

**Summary and conclusions**

In this work, we report two methods, TP&IM and FIB&NM, for the preparation of TEM specimens of QD nanostructures. We demonstrate that our optimization of the combination of FIB&NM has several advantages that can be applied for efficiently preparing high-quality and damage-free TEM specimens consisting of QDs:

a) A FIB lamella offers large thin regions and therefore enhances the probability of finding a complete dot structure in TEM specimens. Thinning of multiple regions effectively reduces the bending of the sample, which is beneficial for TEM characterization.

b) The probability for the presence of QD structures cut through their centers is increased, when cutting the lamella at an angle of 5° with respect to the dot arrays, where the ideal cutting angle is closely correlated to the size and spacing of the dots and needs to be adjusted accordingly.

c) The Fischione NanoMill was used to additionally thin and polish the thick FIB lamellae from both sides. The surface contamination and amorphous layers were gently removed by using $Ar^+$ ions at a low current density, resulting in a flat and clean TEM specimen consisting of SRO QDs on the STO substrate.

With the obtained high-quality TEM specimen, one can carry out more precise structural and chemical studies of QD nanostructures for unraveling the underlying mechanism of their emergent phenomena. The reported settings are practically useful for TEM sample preparations with TP, IM, FIB, and NM. We believe that our optimization of the combined methods (FIB&NM) has significant implications not only for preparing TEM specimens of quantum dots but also for other emerging quantum nanostructures.